\documentclass{iau}
\usepackage{graphicx}
\newcommand{\vect}[1]{\boldsymbol{#1}}

\title[Reconstructing light curves from HXMT imaging observations] %% give here short title %%
{Reconstructing light curves from \\HXMT imaging observations}

\author[Huo \etal]   %% give here short author list %%
{Huo Zhuo-Xi$^1$,
 Zhang Juan$^2$,
 Li Yi-Ming$^1$
 \and Zhou Jian-Feng$^1$}

\affiliation{$^1$Tsinghua University, Beijing, 100084, China \\
email: {\tt huozx@tsinghua.edu.cn} \\[\affilskip]
$^2$Institute of High Energy Physics, Chinese Academy of Sciences,\\
Beijing, 100049, China \\
email: {\tt zhangjuan@ihep.ac.cn}}

\pubyear{2014}
\volume{306}  %% insert here IAU Symposium No.
\pagerange{119--126}
\jname{Statistical Challenges in 21st Century Cosmology}
\editors{A.C. Editor, B.D. Editor \& C.E. Editor, eds.}
\begin{document}

\maketitle

\begin{abstract}
The Hard X-ray Modulation Telescope (HXMT) is a Chinese space telescope mission.
It is scheduled for launch in 2015.
The telescope will perform an all-sky survey in hard X-ray
band ($1$ - $250\;\mathrm{keV}$), a series of deep imaging observations of small
sky regions as well as pointed observations.
In this work we present a conceptual method to reconstruct light curves
from HXMT imaging observation directly, in order to monitor time-varying objects
such as GRB, AXP and SGR in hard X-ray band with HXMT imaging observations.
\keywords{method: data analysis, techniques: image processing, space vehicles,
  X-rays: time-varying objects.}
\end{abstract}

\firstsection % if your document starts with a section,
\section{Introduction}

\subsection{The HXMT mission}

Hard X-ray Modulation Telescope (HXMT) is a planned Chinese space telescope mission.
This is a telescope dedicated to X-ray astronomy in $1$ - $250\;\mathrm{keV}$.
There are three individual telescopes, i.e., the high-energy telescope (HE), the
medium-energy telescope (ME) and the low-energy telescope (LE) on board of HXMT.

HE detects hard X-ray photons from $20\;\mathrm{keV}$ to $250\;\mathrm{keV}$ with
$18$ scintillator detectors.
The field of view (FoV) of each detector is constrained by an individual collimator.
Thus we have $1.1^\circ \times 5.7^\circ$, $5.7^\circ \times 5.7^\circ$ (in FWHM) and
covered FoVs separately.
HE detector can resolve photon-arrival events between
$25\;\mathrm{\mu s}$.
The energy resolution of HE is $19\%$ at $60\;\mathrm{keV}$.
Its effective area is $5\,400\;\mathrm{cm}^2$.

ME covers the energy range from $5\;\mathrm{keV}$ to $30\;\mathrm{keV}$ with $54$
collimated Si-PIN detectors.
It also has two different FoVs, $1^\circ \times 4^\circ$ and $4^\circ \times 4^\circ$
as well.
$24$ collimated CCD detectors are included in LE, which covers the soft X-rays from
$1\;\mathrm{keV}$ to $15\;\mathrm{keV}$.
LE has two different FoVs too.
One is $1.6^\circ \times 6^\circ$ while the other is $4^\circ \times 6^\circ$
(in FWHM).
The effective areas of ME as well as LE are $952\;\mathrm{cm}^2$ and
$384\;\mathrm{cm}^2$ respectively.

There are also a set of anti-coincidence detectors (ACDs) installed around HE
detectors
(\cite[Wu \etal\ 2004]{wu2004})
and a space environment monitor
(\cite[Shen \etal\ 2008]{shen2008})
on board of the satellite to help screen background events and to measure the
spectra and time-varying properties of ambient protons and electrons alongside
the covered HE detector.

To achieve the scientific objectives of this mission the telescope will operate in
three observing modes by its current design.
First, it will perform an all-sky imaging survey in $1$ - $250\;\mathrm{keV}$.
An all sky map in this energy range will be reconstructed from the observed data,
and a catalog of detected objects will be compiled.
At least $25\%$ available observing time of this telescope will be allocated to
the all-sky imaging survey.
Second, it will carry out a series of deep imaging observations of small sky regions.
Maps of these regions will be reconstructed from the observed data.
Finally a considerable percentage of its observing time will be consigned to pointed
observations of various targets to obtain their energy spectra as well as light
curves
(\cite[Lu 2012]{lu2012}).

\subsection{Time-varying objects in hard X-rays}

There are various interesting time-varying objects visible in hard X-rays.
Gamma-ray bursts (GRB) and their X-ray afterglows both contribute to the
their emissions in hard X-rays
(\cite[O'Brien \etal 2006]{obrien2006}).
Because of instability of the accretion disc most compact X-ray sources are
variable
(\cite[Matsuoka \etal 2009]{matsuoka2009}).
Short-term variability of soft Gamma-ray repeaters (SGRs) and anomalous X-ray
pulsars (AXPs) have been detected in hard X-ray.
(\cite[Mereghetti 2008]{mereghetti2008}).

Given its effective area and its FoVs, the geometric factor of HXMT/HE is
competitive in hard X-ray telescopes, which makes it suitable for monitoring
all-sky objects.
It has temporal-resolution as good as $25\;\mathrm{\mu s}$ so it is adequate to
monitor time variability of hard X-ray objects.
However, by its current design HXMT do not distribute burst alert in real time
or perform follow-up observations triggered by other telescopes.
In this work we have present a conceptual pipeline for reconstructing light curves
of time-varying objects solely from HXMT imaging observations.
In this way at least $25\%$ of its total observing time could contribute to
monitoring time-varying objects.

\section{Data analysis pipeline}

\subsection{Temporal-spatial modulation in HXMT imaging observation}
HXMT imaging observation through its collimated detectors is described by
the following temporal-spatial modulation equation
\begin{equation}
d(\vect{\omega}) = 
 \int_t^{t+\Delta t}\int f(t,\vect{x})p(\vect{\omega}(t),\vect{x})
   \mathrm{d}t\mathrm{\vect{x}}
 \approx \Delta t \int f(t,\vect{x})p(\vect{\omega},\vect{x})
   \mathrm{d}\vect{x},
 \label{eq-mod}
\end{equation}
where $\vect{\omega}(t)$ is the status vector of the telescope as a function
of time, $\vect{x}$ is coordinate vector in celestial coordinate system,
$d(\vect{\omega})$ is the observed data as a function of $\vect{\omega}$,
$f(t,\vect{x})$ is the temporal-spatial distribution of the object intensity,
and $p(\vect{\omega},\vect{x})$ is the response of the telescope to a unit
source at $\vect{x}$ while the status of the telescope being $\vect{\omega}$.

\subsection{High-temporal-resolution light curve estimation}
\begin{figure}[htbp]
\centering
\includegraphics[width=0.8\linewidth]{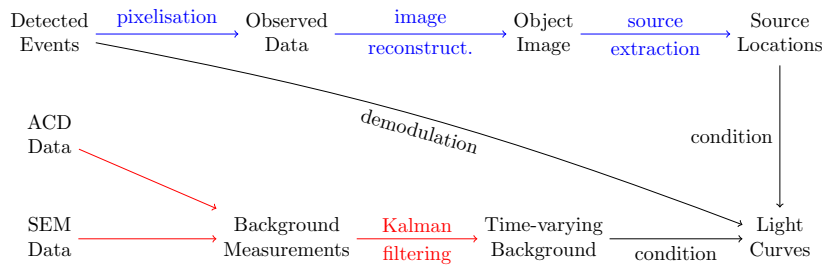}
\caption{Data analysis pipeline}
\label{fig-pipeline}
\end{figure}

To solve Eq. \ref{eq-mod} we present in this work a three-pass pipeline.
The first pass goes through blue paths, while the second and third passes
go through red and black paths in Fig. \ref{fig-pipeline}. 

First, bright sources are located through image reconstruction and source
extraction.
The approximation in Eq. \ref{eq-mod} is valid for time-varying object
only in short time scales.
Hence in each image reconstruction we only use data observed along the same
scanning path.
Therefore the reconstruction problem is of obvious illness, where two-dimensional
image is to be reconstructed from one-dimensional data.
We take advantage of the parallel observations with different collimators and
positions to add on the dimensionality of the observed data.
However it imposes a shift-variant deconvolution difficulty on the image
reconstruction procedure.
The difficulty can be tackled by a clustering-based approximation
(\cite[Huo \& Zhou 2013]{huo2013}).

Second, the time-varying background is estimated by Kalman filtering technique.
The detected events in average originate from background emissions, including
diffusive X-rays, cosmic rays, protons trapped in South Atlantic Anomaly (SAA),
albedo photons, etc.
The diffusive X-rays consist of CXB and Galactic diffuse X-rays.
Cosmic rays have complicated interactions with instruments, which result in a
significant background.
SAA-trapped protons are the dominant factor of HE background.
In addition, this factor will further excite part of the instrument, and induce
a large amount of background after leaving SAA region.
Background induced by albedo photons depends on the elevation and altitude of
the satellite.
Therefore the total background can be summarized as:
\begin{equation}
B_{\mathrm{total}} = B_{\mathrm{CR}}
 + B_{\mathrm{A}}(\vect{\omega}(t))
 + B_{\mathrm{SAA}}(\vect{\omega}(t))
 + B_{\mathrm{G}}
 + B_{\mathrm{C}}
 + B_{\mathrm{I}}
\end{equation}
where $B_{\mathrm{CR}}$, $B_{\mathrm{G}}$, $B_{\mathrm{C}}$ and $B_{\mathrm{I}}$
denote backgrounds from CR, Galactic diffusive X-ray, CXB and the instrument,
which are considered constant by the first-order approximation,
while $B_{\mathrm{A}}(\vect{\omega}(t))$ and
$B_{\mathrm{SAA}}(\vect{\omega}(t))$ represent backgrounds from albedo photons
and SAA-trapped protons, which are time-varying.

Finally, light curves of the objects are demodulated in time domain from
the observed data directly with their locations fixed and the time-varying
background estimated previously.

\section{Summary}

Light curves of the time scales in $10$ - $100\;\mathrm{s}$ or in several days
can be reconstructed from HXMT imaging observations.
We can monitor X-ray time-varying objects bright enough with the conceptual
method suggested in this work, which requires no extra observations.
HXMT is planned to launch in 2015.
Comprehensive feasibility study as well as reliable implementation are in urgent
need amongst all ground-based software researches and developments.

\end{document}